\documentclass[12pt]{article}

\usepackage[utf8]{inputenc}
\usepackage{amsmath,amssymb,bm}
\usepackage[colorlinks=true,linkcolor=blue,citecolor=blue]{hyperref}

\usepackage[symbol]{footmisc} 

\usepackage{physics}
\usepackage{comment}
\usepackage[normalem]{ulem} 

\usepackage{titlesec}
\titleformat*{\section}{\large\bfseries}
\titleformat*{\subsection}{\normalsize\bfseries}

\newcommand{\nn}{\nonumber}
\newcommand{\cl}{\mathcal}

\usepackage[backend=biber, sorting=none, url=false, doi=false, isbn=false,
giveninits=true, uniquename=init]{biblatex}
\begin{document}

\begin{center}
{\LARGE
Recovering the leading quantum correction to Newton Potential beyond the Trace Anomaly}
\vskip 6mm

{\bf Riccardo Fecchio} 
\hspace{-1mm}\footnote{E-mail address: \ riccardo.fecchio@edu.ufes.br},
\vskip 6mm
 PPGCosmo,
 Universidade Federal do Esp\'{\i}rito Santo
\\
 Vit\'oria, 29075-910, ES, Brazil
\vskip 2mm

 Departamento de F\'{\i}sica, ICE,
 Universidade Federal de Juiz de Fora
\\
 Juiz de Fora, 36036-900, Minas Gerais, Brazil

\end{center}

\centerline{{\bf Abstract}}
\begin{quotation}
\noindent
The leading quantum correction to the Newton potential induced by the one-loop vacuum effective action of gravity is of order $1/r^3$.
One-loop corrections are also responsible for the trace anomaly.
We first show, following our previous work, that the effective action obtained by integrating the trace anomaly cannot reproduce this correction. 
We then complete the semiclassical Einstein equations with the vacuum-polarization stress tensor in addition to the anomalous stress tensor.
The full stress tensor follows from the approximation proposed by Visser for the Schwarzschild spacetime. The stress tensor is taken in the Boulware vacuum and expanded at large distance, where it is conserved and reproduces the trace anomaly. Solving the equations perturbatively in $\hbar$, we find that Visser's approximation generates the $1/r^3$ correction to the Newton potential, linear in the source mass $M$, with the same structure as the one-loop result. Its coefficient is controlled by a single free parameter of Visser's polynomial ansatz, which can be tuned to match the one-loop value. The dependence on the trace-anomaly coefficient appears only at order $1/r^4$.
\vskip 3mm

\vskip 3mm


\end{quotation}

\section{Introduction}

General Relativity can be consistently quantized as an effective field theory at energies well below the Planck scale \cite{Donoghue1994,Donoghue1995,Burgess2004}. In this framework the long-distance quantum corrections are universal, in the sense that they do not depend on the unknown ultraviolet completion of quantum gravity. The leading corrections at large distance come from massless particles running in loops, which induce non-local terms that dominate over the local ones. In fact, massive particles in loops contribute only to local terms in the infrared and are thus subleading (see e.g.\ \cite{Shapiro2008} and references therein).

The calculation of the quantum corrections to the Newton potential is the standard illustration of the efficiency of the effective approach \cite{Donoghue:1993eb,Donoghue1994,BjerrumBohr2003}. 
Considering a source of mass $M$ and a test mass $m$, the potential energy obtained from a graviton exchange and the one-loop correction reads
\begin{equation}
V(r) = -\frac{GMm}{r}\left[1 + a\,\frac{G(M+m)}{r} + b\,\frac{G\hbar}{r^2} \right] \,.
\label{eq:intro-pot}
\end{equation}
The first term is the Newton potential, and $a$ and $b$ are numerical coefficients. 
The second term is independent of $\hbar$ and, according to the logic of \cite{Donoghue:1993eb}, this correction is purely classical.
The third term is proportional to $\hbar$ and it is identified the leading quantum correction, of order $1/r^3$. Its value depends on the field content and on the gauge choice \cite{Dalvit1997,BjerrumBohr2003,dePaulaNetto2022}, and it originates from the non-local logarithmic form factors of massless particles, which dominate in the infrared \cite{Donoghue1994,Dalvit1997,Satz:2004hf,dePaulaNetto2022} (see also \cite{Boasso2025} for an analogous result in the case of the horizonless Bardeen metric background).

The logarithmic form factors are also responsible for the trace anomaly \cite{Duff1994,BirrellDaviesBook}. It is therefore natural to ask whether the anomaly alone encodes the leading correction to the Newton potential. A first answer was given in \cite{Donoghue:1993eb}, where it was argued that, because the trace anomaly is local, the $1/r^3$ correction would be missed. 

From the trace anomaly one can extract the so-called anomaly-induced effective action \cite{Riegert1984,FradkinTseytlin1984,Shapiro2008}. It is much simpler than the vacuum effective action derived from explicit one-loop calculations, and it is extremely powerful for situations in which the physics is encoded in the conformal mode of the metric, as in cosmology (see e.g.\ anomaly-induced inflation \cite{Fabris:1998vq}). For the gravitational potential, however, it was shown in \cite{Fecchio2026} (see also \cite{Boasso2025,ABF2005}) that the leading correction is not captured: working with a local version of the anomaly-induced action on a Schwarzschild background, and solving the corresponding semiclassical Einstein equations, the leading quantum correction was found to be of order $1/r^4$ rather than $1/r^3$.

In this paper we complete the semiclassical Einstein equations, sourced by the trace anomaly, with an additional contribution to the stress tensor, called vacuum-polarization, and derive the corresponding Newton potential.
We use the expression worked out by Visser \cite{Visser1997}, which follows from general assumptions \cite{Christensen1977} and from the comparison with approximate results for the quantum stress tensor in semiclassical gravity. We consider it in the Boulware vacuum \cite{Boulware1975}: since we want the potential generated by a static mass, no evaporation has to be taken into account. We also consider a Schwarzschild background, which exploits spherical symmetry and already includes the mass source in the background. The overall stress tensor we use, therefore, describes quantum effects only, namely the trace anomaly and the new terms contained in the vacuum polarization.

The semiclassical Einstein equations are solved perturbatively in $\hbar$, looking for the perturbation $h_{\mu\nu}$ of the Schwarzschild background. This is equivalent to the effective approach used in loop computations \cite{Donoghue1994}, where General Relativity is the only gravitational theory used to derive the leading quantum corrections, linear in $\hbar$.
Since our aim is the gravitational potential, the equations are solved at spatial infinity, with the approximations that this requires.

The main result is that, by completing the trace anomaly with the vacuum polarization in the way suggested in \cite{Visser1997}, the Newton potential acquires a $1/r^3$ term as the leading quantum correction, whose coefficient can be matched to the diagrammatic result \cite{Donoghue1994}.

This result has two consequences. The first is to highlight the quality of Visser’s approximation (indeed, throughout the text we point out the agreement between the approximation and other results obtained in later works). The second is to show that the trace mode of the metric, and equivalently the trace anomaly, does not capture the leading correction to the Newton potential.

This work is organized as follows. In Section~\ref{sec:trace} we review the trace anomaly and show that the trace mode does not carry the $1/r^3$ correction. In Section~\ref{sec:vp} we present Visser's stress tensor in the Boulware vacuum. In Section~\ref{sec:sce} we solve the semiclassical Einstein equations and compare the result with the one-loop correction. Section~\ref{sec:concl} contains our conclusions.

We use the metric signature $(-+++)$ and natural units, writing the constants $G$ and $\hbar$ explicitly when appropriate.

\section{The trace mode and the trace anomaly}
\label{sec:trace}

The one-loop vacuum effective action includes quantum corrections to the background metric. They can be computed via Feynman diagrams or with the more direct heat-kernel technique \cite{BarvinskyVilkovisky1985}. The effective action contains form factors, such as the logarithmic one, which is closely related to the one-loop UV divergences \cite{Buchbinder:2021wzv}.

For the reasons given in the introduction, we focus on massless fields only. The vacuum effective action of such a theory contains logarithmic form factors and reads
\begin{align}
\Gamma = S_{EH} + \bar{\Gamma}
\label{eq:SEH+Gamma}
\end{align}
with
\begin{equation}
\bar{\Gamma} = - \int \dd^4x\sqrt{-g}\left\{\beta_{R^2}\,R\log\!\Big(\frac{-\Box}{\mu^2}\Big)R + \beta_{C^2}\,C_{\mu\nu\alpha\beta}\log\!\Big(\frac{-\Box}{\mu^2}\Big)C^{\mu\nu\alpha\beta}\right\} 
\label{eq:Gamma}
\end{equation}
up to superficial terms and at second order in curvature. Here $\beta_{R^2},\beta_{C^2}$ are the beta functions in the $\overline{\text{MS}}$ scheme, which depend on the number and kind of fields considered, and which reflect the link between the logarithms and the UV divergences.
$R$ and $C_{\mu\nu\alpha\beta}$ are the Ricci scalar and the Weyl tensor, respectively.
Finally,
\begin{equation}
S_{EH} = -\frac{1}{\kappa^2}\int \dd^4x\sqrt{-g}\,R
\label{eq:EH}
\end{equation}
is the Einstein--Hilbert action, with $\kappa^2=16\pi G$.
Let us comment that, in the case of conformal fields, $\beta_{R^2}=0$ for symmetry reasons \cite{Buchbinder:2021wzv}.

The physical effects of the quantum corrections are strongly suppressed by the factor $\hbar$ contained in the beta functions and by the gravitational factor $\kappa^2$. This is illustrated by the derivation of the gravitational potential for the theory \eqref{eq:SEH+Gamma}. In the effective approach, the quantum corrections \eqref{eq:Gamma} are treated as small terms that modify the propagation of a General-Relativistic graviton without introducing new degrees of freedom. The potential is then produced by the exchange of a graviton $h_{\mu\nu}$ between two masses on a flat background $\eta_{\mu\nu}$. It turns out that the logarithmic form factors induce a correction of the form $\hbar G^2M/r^3$ to the Newton potential, times a numerical factor \cite{Dalvit1997,dePaulaNetto2022}.

The logarithms are also responsible for the trace anomaly, because the form factors in \eqref{eq:Gamma} are not conformally invariant. This is seen by a conformal transformation $g_{\mu\nu}=e^{2\sigma}\bar g_{\mu\nu}$, e.g.\ for the Weyl term,
\begin{equation}
\int \dd^4x\sqrt{-g}\,C_{\mu\nu\alpha\beta}\log\!\Big(\frac{-\Box}{\mu^2}\Big)C^{\mu\nu\alpha\beta} = -2\int \dd^4x\sqrt{-\bar g}\,\sigma\,\bar C^{\mu\nu\alpha\beta}\bar C_{\mu\nu\alpha\beta} + f(\sigma,\bar g),
\label{eq:weyl}
\end{equation}
where $f$ is an integral functional that depends on $\sigma$ in a more complicated way than linearly. As a consequence of the conformal breaking, the stress tensor derived from \eqref{eq:Gamma},
\begin{equation}
T^{\alpha\beta} = -\frac{2}{\sqrt{-g}}\frac{\delta\bar{\Gamma} }{\delta g_{\alpha\beta}},
\label{eq:Tdef}
\end{equation}
has a non-zero trace, which is the trace anomaly. Although the stress tensor itself is hard to compute, its trace is not, thanks to the identity
\begin{equation}
-\frac{2}{\sqrt{-g}}\,g_{\mu\nu}\frac{\delta A[g_{\alpha\beta}]}{\delta g_{\mu\nu}} = -\frac{1}{\sqrt{-g}}\,\frac{\delta A[e^{2\sigma}\bar g_{\alpha\beta}]}{\delta\sigma}\bigg|_{\sigma\to0,\ \bar g_{\alpha\beta}\to g_{\alpha\beta}},
\label{eq:identity}
\end{equation}
which holds for any functional $A$ of the metric. The trace obtained from it has the structure
\begin{equation}
T = g_{\alpha\beta} \, T^{\alpha\beta} = -\left(\omega C^2 + bE_4 + c\,\Box R\right) \,,
\label{eq:anomaly}
\end{equation}
with $E_4$ the integrand of the Gauss-Bonnet term, and for some constants $\omega,b,c$, which depend on the beta functions in \eqref{eq:Gamma}. The functionals $f$ that arise from the conformal transformations vanish in the limit $\sigma\to0$, as does the one in \eqref{eq:weyl}, and so do not contribute to the trace anomaly.

The trace anomaly is thus a signature of the one-loop logarithmic corrections. However, it does not carry the same information \cite{Fecchio2026}. If we reverse-engineer the anomaly and solve the functional equation \cite{Riegert1984,FradkinTseytlin1984}
\begin{equation}
-\frac{2}{\sqrt{-g}}\,g_{\mu\nu}\frac{\delta\Gamma_{ind}}{\delta g_{\mu\nu}} = -\left(\omega C^2 + bE_4 + c\,\Box R\right),
\label{eq:induced}
\end{equation}
the resulting anomaly-induced effective action, $\Gamma_{ind}$, contains less information than the original vacuum effective action, for two reasons. First, the solutions of \eqref{eq:induced} are defined up to a conformal action $S_c[g]$, which plays the role of the integration constant of the functional differential equation. Second, the anomaly ``conserves'' the quantum correction of the original theory in the conformal sector only, namely for the conformal mode $\sigma$, while for $\bar g_{\mu\nu}$ the quantum corrections are degraded. 
This implies that the trace anomaly is somewhat inadequate for describing all quantum corrections in phenomena where metric modes other than the trace are relevant.
On the other hand, $\Gamma_{ind}$ is much handier than the action with logarithmic form factors, and it is extremely powerful whenever the relevant physics is encoded in the conformal mode, as in homogeneous and isotropic cosmology, where $\sigma$ is related to the scale factor \cite{Fabris:1998vq}.

In the case of the Newton potential, the conformal mode of the metric, and thus the anomaly-induced effective action, cannot track the leading $1/r^3$ quantum correction. This was studied in detail in \cite{Fecchio2026}, where the local version of $\Gamma_{ind}$ was used to solve the semiclassical Einstein equations on a Schwarzschild background, treating the perturbation $h_{\mu\nu}$ in the effective approach; the leading correction found there was of order $1/r^4$. Here we present an alternative and more direct verification, based on $\Gamma_{ind}$ written with the explicit decomposition $g_{\mu\nu}=e^{2\sigma}\bar g_{\mu\nu}$ \cite{Shapiro2008}:
\begin{align}
\Gamma_{ind} = S_c[g] 
& + \int \dd^4x\sqrt{-\bar g}\left\{\omega\,\sigma \bar C^2 + b\,\sigma\Big(\bar E_4 - \frac23\bar\Box\bar R\Big) + 2b\,\sigma\bar\Delta_4\sigma\right\} \nn\\
& - \frac{1}{12}\Big(c+\frac23 b\Big)\int \dd^4x\sqrt{-g}\,R^2,
\label{eq:Gind}
\end{align}
where $\Delta_4$ is the Paneitz operator \cite{FradkinTseytlin1982a,FradkinTseytlin1982b,Paneitz_2008},
\begin{equation}
\Delta_4 = \Box^2 - 2R^{\mu\nu}\nabla_\mu\nabla_\nu + \frac23 R\,\Box - \frac13(\nabla^\mu R)\nabla_\mu .
\label{eq:paneitz}
\end{equation}
The downside of this formulation is that the action is not covariant. For our purpose this is an acceptable price, for two reasons. First, deriving the gravitational potential requires choosing a static reference frame at spatial infinity, which breaks covariance anyway. Second, an action in which the conformal mode is explicit simplifies the study of its dynamics.

We take the vacuum action to be the Einstein--Hilbert action plus \eqref{eq:Gind}, with a point-like source of mass $M$. We choose a Minkowski background $\bar g_{\mu\nu}=\eta_{\mu\nu}$ and promote $\sigma$ to be the only degree of freedom, namely
\begin{equation}
g_{\mu\nu} = e^{2\sigma}\eta_{\mu\nu}.
\label{eq:conf}
\end{equation}
This assumption is very restrictive, because the Schwarzschild metric does not reduce to \eqref{eq:conf} in the large-radius limit we are interested in. In this limit it reads
\begin{equation}
\gamma_{\mu\nu}\approx \mathrm{diag}\Big(-\Big( 1-\frac{2GM}{r} \Big),\, 1+\frac{2GM}{r},\, r^2,\, r^2\sin^2\theta\Big) \,,
\label{eq:schw}
\end{equation}
and the sign difference between $\gamma_{00}$ and $\gamma_{11}$ shows the incompatibility with \eqref{eq:conf}: a single conformal factor does not cover the degrees of freedom of the Schwarzschild metric.
Or, in other words, the Schwarzschild metric is not a conformally trivial metric.
The appropriate decomposition at large radii would be $g_{\mu\nu} = \eta_{\mu\nu} + h_{\mu\nu}$, with $h_{\mu\nu}$ carrying more than one degree of freedom.
Nevertheless, we proceed with \eqref{eq:conf} to see how the trace anomaly contributes to $\sigma$.

Since our focus is the trace anomaly, we set $S_c=0$. The conformal transformation of the Ricci scalar in four dimensions \cite{Carneiro:2004rt},
\begin{align}
\sqrt{-g}\,R &= \sqrt{-\bar g}\,e^{2\sigma}\Big[\bar R - 6\bar\Box\sigma - 6(\bar\nabla^\alpha\sigma)(\bar\nabla_\alpha\sigma)\Big],\nn\\
\sqrt{-g}\,R^2 &= \sqrt{-\bar g}\Big[\bar R - 6\bar\Box\sigma - 6(\bar\nabla^\alpha\sigma)(\bar\nabla_\alpha\sigma)\Big]^2,
\label{eq:Rconf}
\end{align}
gives the vacuum action for the choice \eqref{eq:conf},
\begin{align}
S_{vac} = S_{EH}+\Gamma_{ind} &= \frac{6}{\kappa^2}\int \dd^4x\, e^{2\sigma}\Big[\Box\sigma + (\partial^\alpha\sigma)(\partial_\alpha\sigma)\Big]\nn\\
&\quad + \int \dd^4x\Big\{2b\,\sigma\Box^2\sigma - 3\Big(c+\frac23 b\Big)\Big[\Box\sigma+(\partial^\alpha\sigma)(\partial_\alpha\sigma)\Big]^2\Big\}.
\label{eq:Svac}
\end{align}
To study the propagation of $\sigma$ we keep the bilinear part of the action; the self-interactions are negligible for $\sigma\ll1$, which is guaranteed for an observer at large distance. The result is
\begin{equation}
S^{(2)}_{vac} = \int \dd^4x\left[\frac{6}{\kappa^2}\,\sigma\Box\sigma - 3c\,\sigma\Box^2\sigma\right].
\label{eq:S2}
\end{equation}
Notice that the $b$-terms coming from the Paneitz operator and from $R^2$ cancel. The action is of fourth order in derivatives, and thus cannot produce a $1/r^3$ term in the potential. For completeness, let us derive the equation of motion. The mass source is described by
\begin{equation}
S_M = -M\int \dd s = -M\int e^{\sigma}\dd\ell,\qquad \dd\ell=\sqrt{\eta_{\alpha\beta}\dd x^\alpha \dd x^\beta},
\label{eq:SM}
\end{equation}
whose contribution to the linearized dynamics of $\sigma$ is
\begin{equation}
S^{(1)}_M = -M\int \sigma\,\dd{\ell} \,.
\label{eq:SM1}
\end{equation}
The equation of motion is
\begin{equation}
\frac{\delta}{\delta\sigma}\Big(S^{(2)}_{vac}+S^{(1)}_M\Big)=0\ \Longrightarrow\ \frac{6}{\kappa^2}\Box\sigma - 3c\,\Box^2\sigma = \frac{M}{2} \,\delta^{(3)}(\mathsf{r}) \,,
\label{eq:eom}
\end{equation}
where $\mathsf{r}$ is the spatial radial vector; the three-dimensional delta survives the integration over $\ell$.
The equation can be solved in Fourier space and, being of fourth order, its solution is expected to have the form of the Stelle potentials, with exponential suppression \cite{Stelle1977}. If instead we follow the effective approach and look for a solution $\sigma=\sigma_c+\sigma_q$, where $\sigma_c$ is sourced by $M$ and $\sigma_q$ by $\hbar$, the exponential suppression is lost, as expected. 
In more detail, at $O(\hbar)$ the equation splits into
\begin{align}
\Box\sigma_c &= \frac{\kappa^2}{12}\,M\,\delta^{(3)}(\mathsf{r}),\label{eq:sc}\\
\Box\sigma_q &= \frac{\kappa^2}{2}\,c\,\Box^2\sigma_c.\label{eq:sq}
\end{align}
The classical solution is
\begin{equation}
\sigma_c(r)=\frac{1}{3}\frac{GM}{r}.
\label{eq:sol-c}
\end{equation}
Combining \eqref{eq:sc} and \eqref{eq:sq} one obtains
\begin{equation}
\Box\sigma_q = \frac{\kappa^4}{24}\,c\,M\,\Box\delta^{(3)}(\mathsf{r}),
\label{eq:sq2}
\end{equation}
and working in Fourier space, after regularizing the small-distance (UV) limit, one finds $\sigma_q=0$.

We conclude that the trace mode $\sigma$ does not carry the leading quantum correction induced by the one-loop vacuum effective action \eqref{eq:SEH+Gamma} as far as the gravitational potential is concerned. This is because the trace mode is not the relevant degree of freedom for the Newton potential.
We have shown this by solving the equation of motion for the trace mode, sourced by a spherically symmetric mass $M$ and by the trace anomaly. 

A more general decomposition, $g_{\mu\nu}=\eta_{\mu\nu}+h_{\mu\nu}$, is instead able to ``feel’’ the leading correction induced by the logarithms of \eqref{eq:SEH+Gamma}, provided that they are not rewritten in terms of the trace anomaly, and gives the leading $1/r^3$ correction to the Newton potential \cite{Dalvit1997,dePaulaNetto2022}. If the logarithms are rewritten in terms of the trace anomaly, the quantum correction ``hits’’ the conformal mode and the resulting correction is subleading \cite{Fecchio2026}. 

Therefore, the trace anomaly must be supplemented with additional contributions in order to properly account for quantum corrections to quantities such as the Newton potential.

\section{Visser's approximation of the stress tensor}
\label{sec:vp}

In this section we set up the element that must be added to the trace anomaly in order to derive a gravitational potential with a $1/r^3$ term.
We work at the level of the semiclassical Einstein equations and find the Newton potential solving the equations in the large radius regime and with the effective approach. 
In \cite{Fecchio2026} this framework was used with the trace anomaly as the only quantum source on the right-hand side of the semiclassical Einstein equations. 
Now, we ``update’’ the quantum source using the stress tensor proposed in \cite{Visser1997} on the right-hand side. 

In general, the quantum stress tensor used in the semiclassical Einstein equations approximates the effects of quantum fields in a vacuum. This quantity remains largely uncertain due to the freedom in choosing the vacuum state of the quantum fields and the difficulty of computing the stress tensor. In fact, only a few analytical results and some numerical approximations are known
(see e.g.\ the seminal work \cite{Christensen1977} and the textbook \cite{Fabbri:2005mw}).

Visser's approximation \cite{Visser1997} is developed on  the Schwarzschild background and reasonably reproduces the known numerical results and satisfies the known analytical constraints on the stress tensor.
We consider Visser's approximation in Boulware vacuum, proposed in Appendix~B of \cite{Visser1997}.

Let us comment that the stress tensor in \cite{Visser1997} is defined with contravariant indices for a local-Lorentz static observer; here, instead, we consider an observer who is static in the usual spherical coordinates in Schwarzschild spacetime.
Because of the spherical symmetry, the full stress tensor is
\begin{align}
T^\mu{}_\nu = 
\begin{pmatrix}
-\rho(r) & +F(r)/f(r) & 0 & 0 \\
-F(r)/f(r)  & -p_r(r) & 0 & 0 \\
0 & 0 & p_t(r) & 0 \\
0 & 0 & 0 & p_t(r) 
\end{pmatrix}
\label{eq:SETfull}
\end{align}
with $f(r) = 1-2GM/r$, and $\rho,p_r,p_t, F$ are respectively the energy density, the radial pressure, the transverse pressure and the net flux.

Let us review the main ideas that led to the approximation proposed in \cite{Visser1997}.

\subsection{Decomposition of the stress tensor}

The starting point is the work of Christensen and Fulling \cite{Christensen1977}, who sought the most general stress tensor in the Schwarzschild spacetime that (a) is conserved, $\nabla_\mu T^{\mu}{}_{\nu}=0$, and (b) reproduces the trace anomaly.
Such a tensor is decomposed into four pieces: 
the piece $T^{ca}_{\mu\nu}$, which reproduces the conformal anomaly in Schwarzschild; the traceless vacuum-polarization piece, $T^{vp}_{\mu\nu}$; and two further traceless pieces describing the energy flux and the momentum density expected for quantum fields in the Unruh and Hartle--Hawking vacua, corresponding to outgoing and ingoing null fluxes, respectively $T^{+}_{\mu\nu}$ and $T^{-}_{\mu\nu}$ \cite{Visser1997,Bardeen2014,Unruh1976,HartleHawking1976}.
Each piece has the same structure as that given in \eqref{eq:SETfull}, although we restric the off-diagonal components $F(r)$ to belong to the flux contribution $T^{+}_{\mu\nu} + T^{-}_{\mu\nu}$ only.

Let us stress that using the decomposition of \cite{Christensen1977} implies the assumption of considering only quantum conformal fields.
Otherwise, there would be no guarantee that the trace of the stress tensor equals the anomalous trace.

In the Boulware vacuum considered here the net flux vanishes, which is equivalent to no evaporation. This does not mean that the flux tensors are identically zero: only the net flux $F(r)$ has to vanish, which leaves room for non-zero diagonal components in $T^{+}_{\mu\nu}+T^{-}_{\mu\nu}$  \cite{Christensen1977,Visser1997,Bardeen2014}.

The anomalous tensor $T^{ca}_{\mu\nu}$ we use here is not exactly the one of \cite{Christensen1977} or \cite{Visser1997}, because we also impose invariance under radial boosts. This assumption is used in the decomposition of \cite{Bardeen2014} instead, and more recently in \cite{Shafiee:2022jfx,Shafiee:2026yuv}.
However, we keep using the notation of \cite{Visser1997}, in which the trace of the stress tensor in Schwarzschild spacetime reads
\begin{equation}
T^{ca}=p_\infty\,\xi\,\frac{(2GM)^6}{r^6} \,, \qquad p_\infty=\frac{\hbar}{90\,\kappa^4(2GM)^4} \,,
\label{eq:Tca}
\end{equation}
where $\xi$ is a number that depends on the conformal quantum fields sourcing the anomaly (see \cite{Visser1997} for a table of values and \cite{BirrellDaviesBook} for the trace anomaly).

The tensor $T^{ca}_{\mu\nu}$ is the easiest contribution to the full stress tensor because its components are fixed by the trace anomaly, the conservation equation and the symmetry of the spacetime.
The tensor is diagonal,  $T^{ca\,\mu}{}_{\nu}=\mathrm{diag}(-\rho^{ca},p^{ca}_r,p^{ca}_t,p^{ca}_t)$, and, because of spherical symmetry, every component depends on $r$ only, and $T^\theta{}_\theta=T^\phi{}_\phi=p^{ca}_t$, while the invariance of the Schwarzschild geometry under radial boosts gives $p^{ca}_r=-\rho^{ca}$. The conservation equation reads
\begin{equation}
\frac{1}{r^2}\Big[p^{ca}_r(-3GM+2r)+GM\rho^{ca}-r(2GM-r)\big(2p^{ca}_t-r\,\partial_r p^{ca}_r\big)\Big]=0,
\label{eq:cons}
\end{equation}
and, together with the trace equation
\begin{equation}
T^{ca}=-\rho^{ca}+p^{ca}_r+2p^{ca}_t,
\label{eq:trace}
\end{equation}
one finds
\begin{equation}
p^{ca}_t=2\rho^{ca},\qquad \rho^{ca}=\frac12\,p_\infty\,\xi\,\frac{(2GM)^6}{r^6}+\frac{c_0}{r^4},
\label{eq:ca-sol}
\end{equation}
where the integration constant $c_0$ is set to zero so that the tensor reproduces the trace anomaly.

Thus, the components of the anomalous stress tensor are fully determined by the trace anomaly. On the other hand, further assumptions are needed to define the vacuum-polarization and flux tensors. 
In \cite{Visser1997}, the components of the stress tensor \eqref{eq:SETfull} are taken to be functionals of the transverse pressure $p_t$ and the trace anomaly (see eqs.\ (7)--(10) of \cite{Visser1997}), while the fluxes are defined by choosing a radiative model.
The functional relations chosen in \cite{Visser1997} are the core of the approximation.
Although we do not report such relations here, since our focus is the final results rather than the intermediate steps, the interested reader can easily find them from the corresponding equation number we cite throughout the text.

The transverse pressure was chosen as a central quantity because several of its properties are known.
Indeed, once an approximation for $p_t$ and a model for the flux are chosen, the components of the stress tensor follow \cite{Visser1997}.

\subsection{The transverse pressure}

To discuss the properties of the transverse pressure $p_t$, we conveniently adopt the dimensionless variable $z = 2GM/r$. 

In the Unruh vacuum, $p_t$ is regular at the horizon \cite{Elster1983,Jensen1991}, whereas the other components of the stress tensor diverge there. Note that, in the decomposition used in this work, the anomalous stress tensor $T^{ca}_{\mu\nu}$ is regular at the horizon for all of its components and for any vacuum state.

At spatial infinity, $z \ll 1$, ref.\ \cite{Christensen1977} argued that in the Unruh vacuum the transverse pressure behaves asymptotically as $\mathcal{O}(z^4)$, as follows from the propagation of massless particles in Schwarzschild spacetime. Since $p_t$ is regular at the horizon and vanishes at infinity, in \cite{Visser1997} a polynomial approximation for the transverse pressure in the Unruh vacuum is proposed: $\sum a_n z^n$ with $n \geq 4$.

We now turn to the Boulware vacuum. Near the horizon, $z \sim 1$, the stress tensor in this vacuum is known to diverge \cite{Christensen1977}, while at infinity it must vanish in order to recover Minkowski spacetime.
Ref.\ \cite{Christensen1977} proposed an asymptotic behavior of $\mathcal{O}(z^6)$ at spatial infinity, based on the idea that, if nothing but the anomaly were present, the behavior would be $z^6$, as it is for the anomaly itself.

In \cite{Visser1997}, a polynomial approximation for the transverse pressure is adopted also for the Boulware vacuum, as done for the Unruh vacuum. 
The idea is to use this approximation from infinity down to some distance $z < 1$ outside the horizon. Near the horizon, the approximation breaks down because the polynomial does not diverge at $z=1$. 
Importantly, in the Boulware vacuum, the asymptotic behavior of the transverse pressure is modified to $\mathcal{O}(z^3)$. 
To determine it, \cite{Visser1997} required the full stress tensor to vanish at $z = 0$. This requires the sum of the anomalous, vacuum polarization, and flux parts (the latter being only diagonal in Boulware) to vanish at infinity. Since the anomalous piece is subleading, going as $\mathcal{O}(z^6)$, this condition relates the flux $T^{+}_{\mu\nu}+T^{-}_{\mu\nu}$ and the vacuum polarization $T^{vp}_{\mu\nu}$ at infinity.
Due to the initial assumption of defining the vacuum polarization components as functionals of the transverse pressure, the condition for recovering Minkowski at $z = 0$ allows to write the flux as a function of the transverse pressure. 
As a result, it was found that $\rho$ and $p_r$ vanish at infinity if $p_t \sim \mathcal{O}(z^3)$ there (see Appendix~B of \cite{Visser1997}).

Therefore, $p_t \sim \mathcal{O}(z^3)$ at infinity is compatible with the recovery of Minkowski spacetime, and in Visser's approximation of the stress tensor in the Boulware vacuum, the transverse pressure is a polynomial series starting at order $z^3$.

\subsection{Components in the Boulware vacuum}

The stress tensor developed in \cite{Visser1997}, in the Boulware vacuum, is thus diagonal (no fluxes), and determined by the polynomial ansatz of the transverse pressure and the trace anomaly only.
In components,
\begin{align}
\rho(z) &= p_\infty\left[-k_3\frac{z^4}{2(1-z)}-\xi\,\frac{z^6(10-9z)}{10(1-z)}+S_k(z)-S_{kk}(z)\right],\label{eq:rho}\\
p_r(z) &= p_\infty\left[k_3\frac{z^3(4-3z)}{1-z}+\xi\,\frac{z^7}{10(1-z)}+S_k(z)+S_{kk}(z)\right],\label{eq:pr}\\
p_t(z) &= p_\infty\left[k_3 z^3+S_k(z)\right],\label{eq:pt}
\end{align}
where we defined the series
\begin{align}
S_k(z) &= \sum_{n=4}^{\infty}k_n z^n,\label{eq:Sk}\\
S_{kk}(z) &= \frac{z^2}{1-z}\sum_{n=5}^{\infty}(n-4)\,k_n\left(\frac{z^{n-1}}{n-1}-\frac{z^{n-2}}{n-2}\right).\label{eq:Skk}
\end{align}
The $k_n$ are dimensionless free parameters of the polynomial ansatz for the transverse pressure: they are not fixed by conservation and the trace anomaly alone (all but one, as we are going to argue), and this freedom is exploited below. The formulas above describe the full stress tensor in Boulware vacuum outside the horizon, for $z < 1$.

The $z^3$ term of $p_t$ is allowed by the mathematics but uncomfortable. For all the components to be of the same order at spatial infinity ($z\ll1$) we must set $k_3=0$. This choice is also needed to respect the anomalous trace and the conservation: for $z\ll1$ the trace of the full stress tensor reduces to the anomalous one only if $k_3=0$, giving
\begin{equation}
T=p_\infty\,\xi\,\frac{(2GM)^6}{r^6}+O\!\left(1/r^7\right),
\label{eq:Tasym}
\end{equation}
Likewise, $\nabla_\mu T^{\mu}{}_{\nu}=0$ is achieved only with $k_3=0$.
The other coefficients might be constrained by fitting the numerical approximations of the stress tensor, as e.g.\ done in the Unruh vacuum state \cite{Visser1997}.

\section{Semiclassical Einstein equations}
\label{sec:sce}

We use the stress tensor above as the source of the semiclassical Einstein equations,
\begin{equation}
G_{\mu\nu}=\frac{\kappa^2}{2}\,T_{\mu\nu} \,,
\label{eq:sce}
\end{equation}
to find the backreaction on the Schwarzschild background metric. 
Although the equations \eqref{eq:sce} seem to be the usual Einstein equations, the stress tensor on the right-hand side is linear in $\hbar$, through $p_\infty$, and introduces one-loop effects in the geometry. Furthermore, we recall that these effects are evaluated for quantum fields in the Boulware vacuum. 

The semiclassical Einstein equations are a pragmatic approach, where we do not care which action $T_{\mu\nu}$ comes from, and simply introduce it as a source. This differs from \cite{Dalvit1997,dePaulaNetto2022}, where the starting point is the one-loop vacuum effective action. 
In this approach it is also unnecessary to specify the vacuum state. 

We expect the two approaches to give the same $1/r^3$ correction to the Newton potential once the appropriate stress tensor $T_{\mu\nu}$ is considered.

The Newton potential is related to the $00$ component of the metric, evaluated at large distance. So we solve \eqref{eq:sce} asymptotically, in agreement with the approximation of the stress tensor, which follows from the polynomial ansatz for the transverse pressure. We also employ the effective approach: $g_{\mu\nu}=\gamma_{\mu\nu}+h_{\mu\nu}$, where $\gamma_{\mu\nu}$ is the Schwarzschild metric and $h_{\mu\nu}$ is the quantum correction sourced by $T_{\mu\nu}$. The equations are expanded perturbatively in $\hbar$, and the asymptotic behaviour is imposed by a polynomial ansatz for $h_{\mu\nu}$.

At zeroth order in $\hbar$ the equation is an identity for the Schwarzschild metric,
\begin{equation}
G_{\mu\nu}^{(0)} (\gamma; x) = R_{\mu\nu}(\gamma; x) -\frac12 \gamma_{\mu\nu} R(\gamma; x) = 0 \,.
\end{equation}
The first-order equation is obtained by expanding the Einstein tensor,
\begin{equation}
G^{(1)}_{\mu\nu}(\gamma,h;x)=\int \dd^4y\,\frac{\delta G_{\mu\nu}(\gamma;x)}{\delta\gamma_{\alpha\beta}(y)}\,h_{\alpha\beta}(y)=\frac{\kappa^2}{2}\,T_{\mu\nu}(\gamma;x) \,.
\label{eq:G1}
\end{equation}
In a generic background $g_{\mu\nu}$, a direct calculation gives
\begin{align}
G^{(1)}_{\mu\nu}(g,h)=\frac12\Big[&\nabla_\lambda\nabla_\mu h^\lambda{}_\nu+\nabla_\lambda\nabla_\nu h^\lambda{}_\mu-\nabla_\mu\nabla_\nu h-\Box h_{\mu\nu} \nn\\
&-g_{\mu\nu}\big(\nabla_\alpha\nabla_\beta h^{\alpha\beta}-\Box h-R_{\alpha\beta}h^{\alpha\beta}\big)-R\,h_{\mu\nu}\Big],
\label{eq:G1gen}
\end{align}
where the covariant derivatives and $\Box=g^{\mu\nu}\nabla_\mu\nabla_\nu$ are built with the background metric, which is also used to raise and lower indices. On the Schwarzschild background, which is Ricci flat, this reduces to
\begin{equation}
G^{(1)}_{\mu\nu}(\gamma,h)=\frac12\Big[\nabla_\lambda\nabla_\mu h^\lambda{}_\nu+\nabla_\lambda\nabla_\nu h^\lambda{}_\mu-\nabla_\mu\nabla_\nu h-\Box h_{\mu\nu}-\gamma_{\mu\nu}\big(\nabla_\alpha\nabla_\beta h^{\alpha\beta}-\Box h\big)\Big].
\label{eq:G1schw}
\end{equation}

\subsection{Asymptotic expansion and solution}

By spherical symmetry, the perturbation depends on two functions of the radial coordinate, $u(r)$ and $v(r)$:
\begin{equation}
h_{\mu\nu}=\mathrm{diag}\Big(-2u(r),\,2v(r),\,2v(r)\,r^2,\,2v(r)\,r^2\sin^2\theta\Big) \,.
\label{eq:hansatz}
\end{equation}
We approximate them by the asymptotic polynomials
\begin{align}
u(r) = \sum_{n=2}^4 \frac{u_n}{r^n} \,, \qquad
v(r) = \sum_{n=2}^4 \frac{v_n}{r^n} \,,
\label{eq:uvseries}
\end{align}
with dimensionful constants $u_n, v_n$. The series starts at $n=2$ because, for $z\ll1$ and $k_3=0$, the stress tensor is $\cl{O}(z^4)$, and since the Einstein tensor is of second order in derivatives, the asymptotic solution for $h_{\mu\nu}$ is $\cl{O}(1/r^2)$. We stop at $n=4$, which means sourcing $h_{\mu\nu}$ with the stress tensor up to $\cl{O}(1/r^6)$, the order of the trace anomaly, one of the assumptions used to build the tensor. Higher orders could be considered, but they go beyond the purpose of recovering the $1/r^3$ one-loop correction.

Inserting the ansatz \eqref{eq:hansatz} into the first-order expansion of the Einstein tensor, \eqref{eq:G1schw}, fully determines the left-hand side of the semiclassical Einstein equations \eqref{eq:G1}. 

Let us now turn our attention to the right-hand side. 
The expansion of the full stress tensor's components for $z\ll1$, with $k_3=0$, up to $\cl{O}(1/r^6)$, is
\begin{align}
\rho &= p_\infty k_4 \qty(\frac{2GM}{r})^4 +\frac43 p_\infty k_5 \qty(\frac{2GM}{r})^5 +\frac{1}{12} p_\infty (k_5 + 18 k_6 -12 \xi) \qty(\frac{2GM}{r})^6 \,, \label{eq:rholr}\\
p_r &= p_\infty k_4 \qty(\frac{2GM}{r})^4 + \frac23 p_\infty k_5 \qty(\frac{2GM}{r})^5
-\frac{1}{12} p_\infty (k_5 - 6 k_6) \qty(\frac{2GM}{r})^6 \,, \label{eq:prlr}\\
p_t &= p_\infty k_4 \qty(\frac{2GM}{r})^4 + p_\infty k_5 \qty(\frac{2GM}{r})^5 + p_\infty k_6 \qty(\frac{2GM}{r})^6 \,. \label{eq:ptlr}
\end{align}
We note that the relations among the components of the stress tensor of \cite{Visser1997}, eqs.\ \eqref{eq:rholr}--\eqref{eq:ptlr}, are in excellent agreement with the numerical results for the stress tensor of a massless scalar field with arbitrary coupling to the scalar curvature in the Boulware state on Schwarzschild spacetime \cite{ABF2005}, once the conformal limit is taken, and provided that we set $k_4=0$. This agreement provides an initial indication that $k_4$ may be problematic. We will retain it for the time being and return to this issue later, explaining why setting $k_4=0$ is a well-motivated choice.

Inserting \eqref{eq:hansatz}--\eqref{eq:uvseries} into the left-hand side of \eqref{eq:G1}, and \eqref{eq:rholr}--\eqref{eq:ptlr} into the right-hand side, the semiclassical Einstein equations reduce to an algebraic system for $u_n$ and $v_n$, with solutions
\begin{align}
& u_2 = \frac{\hbar G}{5760\pi} k_4 \,, \qquad \qquad u_3 = -\frac{\hbar G^2M}{103680\pi} (9k_4-16k_5)\,, \nn \\
& \qquad u_4 = -\frac{\hbar G^3M^2}{103680\pi} \big(15k_4+5k_5-6(3k_6-\xi)\big) \,,
\label{eq:usol}
\end{align}
and
\begin{align}
& v_2 = -\frac{\hbar G}{11520\pi} k_4 \,, \qquad \qquad
v_3 = - \frac{\hbar G^2M}{103680\pi} (15k_4+8k_5) \,, \nn \\
& \qquad v_4 = - \frac{\hbar G^3M^2}{414720\pi} \big(69k_4+58k_5+12(3k_6-2\xi)\big) \,.
\label{eq:vsol}
\end{align}
In presenting these solutions we substituted both $\kappa^2=16 \pi G$ and $p_\infty$ from \eqref{eq:Tca}.

\subsection{The Newton potential and the matching with the one-loop correction}

The Newtonian potential $\Phi$ is read from the time-time component of the metric, $g_{00}=\gamma_{00}+h_{00}=-(1+2\Phi)$, which gives, using our large radius solution,
\begin{align}
\Phi(r) &= -\frac{GM}{r} + u(r)  \nn \\
&= - \frac{G M}{r} + \frac{\hbar G}{5760\pi} k_4 \frac{1}{r^2}
- \frac{\hbar G^2M}{103680\pi} (9k_4-16k_5)\frac{1}{r^3}  \nn \\
& \qquad \qquad -\frac{\hbar G^3M^2}{103680\pi}
 \big(15k_4+5k_5-6(3k_6-\xi)\big) \frac{1}{r^4} \,.
\label{eq:Phi}
\end{align}
The $1/r^2$ quantum correction can be switched off by setting $k_4=0$. This is the natural choice: that term is proportional to $\hbar$ but independent of $M$, so it cannot be a correction to the potential generated by the mass; it is a spurious effect of the approximation of the stress tensor. 
Furthermore, in the logic of \cite{Donoghue:1993eb}, the only expected $1/r^2$ correction is the relativistic one, which is of order $\hbar^0$. With $k_4=0$ the $1/r^3$ order becomes the first quantum correction,
\begin{equation}
\Phi(r)=-\frac{GM}{r} + \frac{\hbar G^2 M }{6480\,\pi}\,\frac{k_5}{r^3}+O\!\left(\frac{1}{r^4}\right),
\label{eq:Phi3}
\end{equation}
and the metric functions satisfy $u_3=-2v_3$. Dividing \eqref{eq:intro-pot} by the test mass, $\Phi=V/m$, the one-loop correction is $-b\,G^2M\hbar/r^3$. The two corrections, \eqref{eq:intro-pot} and \eqref{eq:Phi3},  have the same functional form: they are linear in $M$ and proportional to $G^2\hbar$. They coincide if
\begin{equation}
k_5 = - 6480\,\pi\, b \,.
\label{eq:matching}
\end{equation}
Thus the free parameter $k_5$ can be tuned to reproduce the one-loop diagram result.

It is worth stressing the meaning of this result. The parameter $k_5$ is a coefficient of the polynomial ansatz for the transverse pressure and is not determined by conservation and the trace anomaly; the matching \eqref{eq:matching} fixes this freedom rather than predicting the number $b$. What is non-trivial is the structure of the outcome. First, a $1/r^3$ correction is generated, which the trace anomaly alone could not produce, and is linear in $M$. 
Second, the relation $u_3=-2v_3$ between the two metric functions is fixed.
This relation also appears in \cite{dePaulaNetto2022}; see eqs.\ (57)--(58) for the conformal case, $\beta_0=0$, and in the absence of interactions between the quantum conformal matter and the classical mass sources, i.e. $\beta_{RT1}=\beta_{RT2}=\beta_{TT}=0$.
Finally, the anomaly coefficient $\xi$ appears only at $\cl{O}(1/r^4)$ in \eqref{eq:usol}--\eqref{eq:vsol}. This is consistent with \cite{Fecchio2026}, where the leading correction obtained from the anomaly-induced effective action was of order $1/r^4$.

\section{Conclusions}
\label{sec:concl}

We have shown that the trace anomaly alone does not account for the leading quantum correction to the Newton potential. In a decomposition $g_{\mu\nu}=e^{2\sigma}\eta_{\mu\nu}$ the conformal mode obeys a fourth-order differential equation, derived form the vacuum action obtained by supplementing the Einstein-Hilbert action with the anomaly-induced effective action. 
In the effective approach, the quantum part of the trace mode vanishes. This is in qualitative agreement with \cite{Boasso2025,Fecchio2026,ABF2005}. Explicitely, in \cite{Fecchio2026} the leading correction from the anomaly-induced action was derived by solving the semiclassical Einstein equations and was found to be of the subleading order $1/r^4$.

To fill this gap, we completed the semiclassical Einstein equations with Visser's approximation of the stress tensor in the Boulware vacuum \cite{Visser1997}, which is conserved and has the anomalous trace.
The equations are then considered at large distance, expanding the stress tensor on the right-hand side up to $\cl{O}(1/r^6)$, the order of the trace anomaly in Schwarzschild. 
The effective approach is implemented by solving the equations perturbatively in $\hbar$ for a metric decomposition $g_{\mu\nu} = \gamma_{\mu\nu}+h_{\mu\nu}$.  
Such decomposition is fundamental because of the stress tensor formulated in \cite{Visser1997} has the anomalous trace, and thus the trace mode of the metric perturbation does not capture the leading correction for the Newton potential.

In the end, we obtained a $1/r^3$ correction to the Newton potential, linear in the source mass, of the same form as the one-loop result. Its coefficient is controlled by the single parameter $k_5$ of Visser's polynomial ansatz for the transverse pressure, used for building the stress tensor \cite{Visser1997}.
Also, the dependence on the anomaly coefficient $\xi$ first appears at order $1/r^4$, as previously verified \cite{Fecchio2026}.

Our result shows that the one-loop $1/r^3$ term can be accommodated by a stress tensor that is conserved and has an anomalous trace \cite{Visser1997}, and that reproduces known relations among its components and among the metric corrections it sources \cite{dePaulaNetto2022, ABF2005}.
Furthermore, the coefficient of the $1/r^3$ term obtained from the diagrammatic computation can be reproduced by tuning the free parameter $k_5$ in the approximation of the stress tensor, see eq.\ \eqref{eq:matching}.

\section*{Acknowledgements}

The author is grateful to I.L.\ Shapiro for many discussions over the past years, and to Centro Brasileiro de Pesquisas Físicas (CBPF) for the hospitality during which this manuscript was finalised.
This work is supported by Coordenação de Aperfeiçoamento de Pessoal de Nível Superior – CAPES (Brazil).

\printbibliography

\end{document}